\documentclass[%
 reprint,
groupedaddress,
 amsmath,amssymb,
aps,
pra,
floatfix,
]{revtex4-1}

\usepackage{graphicx}
\usepackage{dcolumn}
\usepackage{bm, bbm}
\usepackage{xcolor}

\usepackage{comment}

\begin{document}

\preprint{Flav-vw}

\title{No more gap-shifting: Stochastic many-body-theory based TDHF for accurate theory of polymethine cyanine dyes}

\author{Nadine C. Bradbury}
\author{Barry Y. Li.}
\author{Tucker Allen}
\author{Justin R. Caram}
\email{jcaram@chem.ucla.edu}
\author{Daniel Neuhauser} 
\email{dxn@ucla.edu}
\affiliation{Department of Chemistry and Biochemistry, University of California, Los Angeles, CA, 90095, USA}


\date{\today}

\begin{abstract}
We introduce an individually fitted screened-exchange interaction for the time-dependent Hartree-Fock (TDHF) method and show that it resolves the missing binding energies in polymethine organic dye molecules compared to time-dependent density functional theory (TDDFT). The interaction kernel, which can be thought as a dielectric function, is generated by stochastic fitting to the screened-Coulomb interaction of many-body perturbation theory (MBPT), specific to each system. We test our method on the flavylium (Flav) and indocyanine green (ICG) dye families with a modifiable length of the polymethine bridge, leading to excitations ranging from the visible to short-wave infrared (SWIR). Our approach validates earlier observations on the importance of inclusion of medium range exchange for the exciton binding energy. Our resulting method, TDHF@$v_W$, also achieves a mean absolute error on par with MBPT at a computational cost on par with local-functional TDDFT.
\end{abstract}

\maketitle



In 1948, Hans Kuhn showed that the optical transitions of organic cyanine dyes could reasonably be described by the ``particle in a box" (PIB) free electron gas model.\cite{Kuhn1948} Such a simplistic quantum theory works astoundingly well for these systems. In fact, since then it has been widely shown that the commonly used modern quantum chemistry methods for optical spectroscopy, time-dependent Hartree-Fock (TDHF), and all varieties of time-dependent density functional theory (TDDFT), have much lower accuracy, often up to 1.0 eV error in transition energy.\cite{LeGuennic2015} This is known as the ``cyanine problem" of TDDFT. 

The PIB model succeeds because the PIB orbitals do resemble the frontier molecular orbitals (MO's) of cyanine dyes, with both the highest-occupied $\pi$ MO (HOMO) and the lowest-unoccupied $\pi^*$ MO (LUMO) fully delocalized over the large polymethine backbone. With modern quantum chemistry methods, achieving such full delocalization is challenging without a full long-range exchange interaction, rendering local DFT methods poorly suited. The complete exchange interaction of HF may yield accurate-looking molecular orbitals, but correct excitation energies require a detailed balance of long-range charge transfer and local valence excitations. Indeed, the amount of HOMO to LUMO wavefunction overlap and the predicted excitation energy have been shown to be directly proportional to the medium range component of the exchange in a chosen hybrid density functional.\cite{Jacquemin2007} Given such high dependency on functional choice in these systems, it is common to apply an arbitrary rigid energy shift when comparing to experimental spectra.

To provide an $ab$ $initio$ approach to correct excitation energies, many-body perturbation theory (MBPT) methods, such as the $GW$+Bethe-Salpeter Equation (BSE) approach, have been shown to achieve much better accuracy ($\sim$ 0.1 eV).\cite{Boulanger2014,Jacquemin2017} This improved accuracy comes at a high cost, as these methods scale steeply with system size, so they are very expensive to be applied for large molecules with hundreds of electrons as used in biological imaging with very low excitation energies in the near and short-wave infrared range (NIR: $\lambda=700 - 1000$ nm and SWIR: $\lambda=1000 - 2000$ nm).

Recently, we have developed stochastic methodologies that reduce the computational scaling of the BSE, unlocking the study of systems up to several thousands of electrons.\cite{bradbury_bethesalpeter_2022,bradbury_optimized_2023} In these works, we pair an iterative approach to calculate the full optical spectra of the molecule, with a stochastic approach to determine the screened-Coulumb interaction $W$, the typically expensive part of the BSE.\cite{bradbury_bethesalpeter_2022} In our most recent work, we use stochastic time-dependent Hartree (TDH) propagation to fit a transitionally invariant exchange kernel $v_W(r-r')$ which is custom for each system and has a nearly equivalent effect acting on a pair density as the full screened-Coulumb interaction $W(r,r')$ and similar overall accuracy for the optical gap.\cite{bradbury_optimized_2023} We term this individually fitted interaction $v_W$, and when replacing the full $W$ in BSE with $v_W$, the resulting equation is a version of TDHF using $v_W$ in the exchange response kernel, henceforth refered to as TDHF@$v_W$.

Building a modified exchange kernel is the original principle behind the optimally tuned Fock exchange contribution in range-separated hybrid functional methods, where the smoothness and the total amount of long-range exchange are fitted to give accurate ionization potentials by partial charging of the system.\cite{Baer2010rsh} However, for many systems, limiting the functional form of the exchange kernel is too strong of an assumption. Instead of imposing a specific functional form on the exchange, we directly access the exchange kernel, or dielectric function, at all length scales using stochastic fitting.

\begin{figure*}
\includegraphics[width=6.5 in]{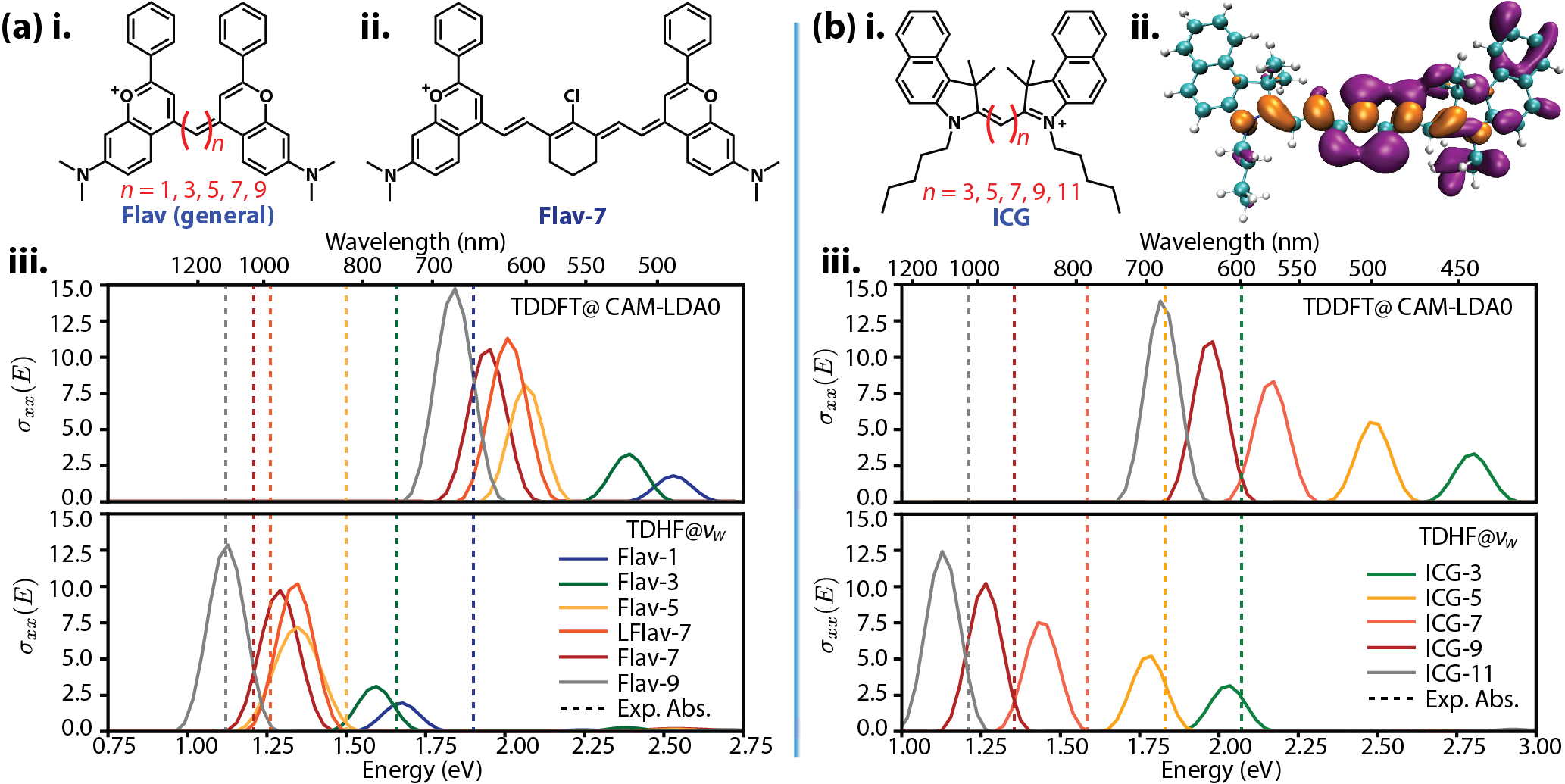}
\caption{\label{fig:wide} (a) Lewis structures (i,ii) and linear response absorption cross section (iii) for the Flav dye family for TDDFT@CAM-LDA0 and TDHF@$v_W$. Experimental $\lambda_{\text{max}}$ in water solution (dashed vertical lines) for the Flav dyes are extracted from Ref.\cite{Cosco2017_ange}. (b) Lewis structure (i), ICG-7 exciton density showing the elctron (purple) and hole (orange) (ii), and linear response absorption cross section (iii) for the ICG dye family using both TDDFT@CAM-LDA0\cite{Mosquera2016} and TDHF@$v_W$. Experimental $\lambda_{\text{max}}$ in water solution values for the ICG dyes are extracted from Refs.\cite{Swamy2024,Langhals2011,Gamage2024,Heng2022}. }
\end{figure*}

\begin{figure*}
\includegraphics[width=6.5 in]{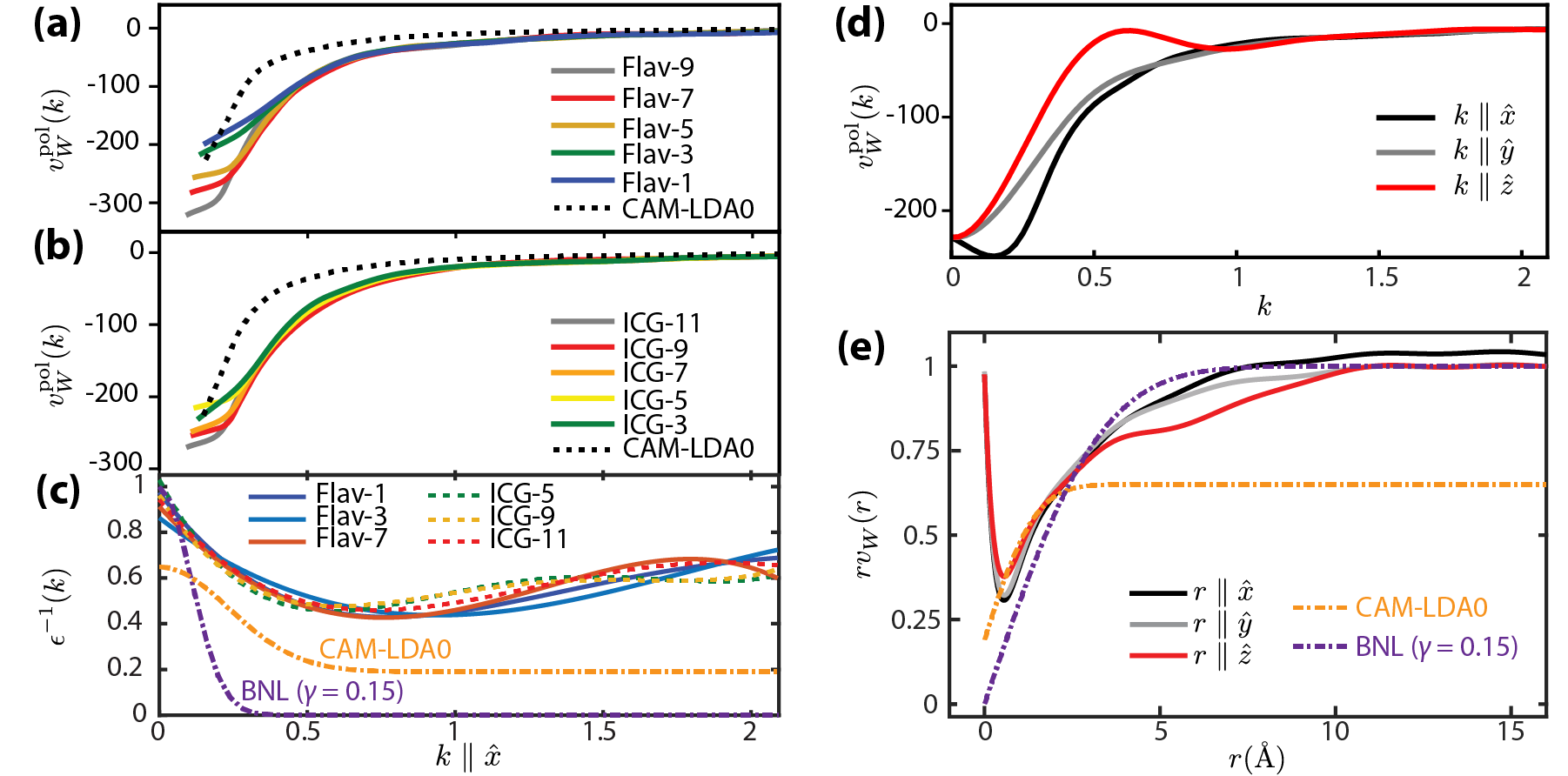}
\caption{\label{fig:vw} Plot of individually-fitted $v_W(k)$ along the $x$-axis of the polymethine bridge for the (a) Flav and (b) ICG dye families. All $v_W $ shown are generated by fitting with 2000 stochastic orbitals. (c) shows the shifted ratio between the fitted $v_W(k)$ and the bare Coulomb potential $v(k)=4\pi/k^2$, i.e., $\epsilon^{-1}(k)\equiv1+k^2v_W(k)/(4\pi)$.  (d) The spatial components of the fitted $v_W^{\rm{pol}}$ along the cardinal $k$-paths $(k_x,0,0)$, $(0,k_y,0)$, and $(k_z,0,0)$, and the complement in real space (e) showing $rv_W(r)$, an alternative representation of the dielectric screening at varying inter-atomic distances. All variables are in atomic units unless otherwise stated.}
\end{figure*}

While TDHF is less expensive than the full BSE, it can still be considerably costly for larger systems. To reduce the effective scaling, we use a split deterministic-sparse stochastic compression scheme to efficiently calculate and store all the attenuated exchange integrals.\cite{bradbury_neargap_2023,Sereda2024} In this work, we study the flavylium (Flav) and indocyanine green (ICG) family of dyes due to their tunable and experimentally well-characterized molecular structures, bulkiness in size, and outstanding impact in the fields of shortwave infrared (SWIR) used for biological imaging.\cite{Cosco2017_ange,Cosco2020_naturechem,Alander2012} All the dye geometries are optimized at the PBE0/def2-TZVPP level of theory using the ORCA 5.0 program.\cite{SeeSI,Neese2022} We show that a modified exchange kernel, which includes more exchange at ``medium-$k$" than conventional range-separated hybrid functionals, recovers the missing binding energies found in these dye molecules.

The primary object of theoretical interest in this letter is the translationally invariant screened-Coulomb interaction in the style of MBPT. The theoretical development of the optimization of this interaction is detailed in Ref. \cite{bradbury_optimized_2023} and summarized here. The translationally invariant screened potential is a function of the form $v_W(r-r')$ that minimizes the difference from the true static $W(r,r')$ screened potential over all occupied orbital pair densities $(\phi_i\phi_j)$. Instead of sampling all occupied densities to construct each screened exchange kernel, we introduce a set of stochastic pair density samples $\bar \beta(r) = \sum_i(\pm1)\phi_i(r)$, where $i\leq N_{\text{occ}}$. Since we choose $v_W(r-r')$ to be  diagonal in momentum space, the optimal form of $v_W$ is extracted as 
\begin{equation}
   \label{eq:vw_opt}
   v_W(k) =
   \frac{\Big\{ \beta^*(k)\langle k |W|\beta\rangle \Big\} }
   {\Big\{  \big|\langle k |\beta \rangle \big|^2 \Big\} }.
\end{equation}
where $\{...\}$ indicates a statistical average, and $\beta(r)\equiv \bar{\beta}(r)\bar{\bar{\beta}}(r)$ is a pair density of independent stochastic orbitals. 

The crucial ingredient in Eq.\ref{eq:vw_opt} is the action of the full many-body $W$ on the random pair density, $\langle k |W|\beta\rangle = W_\beta(k)$. This is obtained as introduced in Refs. \cite{bradbury_optimized_2023,bradbury_bethesalpeter_2022,Vlek20181} through a stochastic TDH propagation with a source potential derived from $\beta(r)$, and summarized here. In MBPT, $W = v + v\chi v \equiv v + W^{\rm{pol}}$ is derived from a static Coulomb interaction and a polarization-mediated interaction component.\cite{Blase2020} The action can be written as $W_\beta = q_\beta + W_\beta^{\rm{pol}}$ with source potential $q_\beta = \int dr' \beta(r')|r-r'|^{-1}$. The polarization component is generated from a TDH propagation comparing unperturbed and perturbed stochastic orbitals, $\eta'_s = e^{i\alpha q_\beta(r)}\eta_s(r)$, evolved by
 \begin{equation}
     i\dot \eta'_s(r,t) = (H_0 + u_\beta(r) + \hat X_0 (t))\eta'_s(r,t),
 \end{equation}
 where $u_\beta(r,t) = \int dr' |r-r'|^{-1}(n^\alpha(r,t) - n^{\alpha=0}(r,t))$, made from a stochastic density $n^\alpha(r,t) = 2\{ |\eta_s(r,t)|^2\}_s$. Next, the exchange term is completed with stochastic orbitals, as recently developed for $GW$ theory in Ref. \cite{Allen2024}. Using a hybrid-orbital starting point for this step in MBPT has been shown to improve results substantially.\cite{McKeon2022}
 
 Another crucial aspect of achieving good results in stochastic propagation for the action is the orthogonality routine, which removes the occupied-occupied contamination in the propagated perturbed molecular orbitals.\cite{bradbury_optimized_2023} Specifically, approximately every dozen time steps, a projection of the stochastic orbitals is performed to remove this contamination, i.e.,
 \begin{equation}
     \eta_s^\alpha(t) \to \eta_s^{\alpha=0}(t)-(\bm{I}-\bm{P})(\eta^\alpha(t) - \eta^{\alpha=0}(t)),
 \end{equation}
with $\bm{P}$ being a projection onto all occupied orbitals, followed by a renormalization of $\eta_s$.
 
 Finally, the action of $W$ on $\beta$ is obtained by
 \begin{equation}
     W^{pol}_\beta(r) = \alpha^{-1} \int dt u_\beta(r,t) e^{-\gamma^2t^2},
 \end{equation}
where $\gamma = 0.1$ a.u. serves as a numerical damping parameter for efficient convergence and the statistical error is sufficiently controlled even when only 10 stochastic $\eta$ orbitals are used. 

One can consider that $v_W^{\rm{pol}} = v_W - v$ contains information from the dielectric function of the molecule, following the expression $1 + \epsilon^{-1}(k) = v_W^{\rm{pol}}(k)/v(k)$, with $v$ the bare non-periodic Coulomb interaction.

Optical electronic absorption spectra for these chosen dyes are then extracted using a TDHF simulation with the fitted $v_W$ made uniquely for each dye molecule. The full technique for this calculation is detailed in Ref. \cite{Sereda2024}, where it is benchmarked to other electronic structure software. For each dye molecule, we propagate all occupied orbitals and $400$ unoccupied orbitals, and use a total of 1000 short stochastic fragments in the compression of the exchange kernel.\cite{SeeSI}

\begin{figure}[b]
\includegraphics[width=3.24 in]{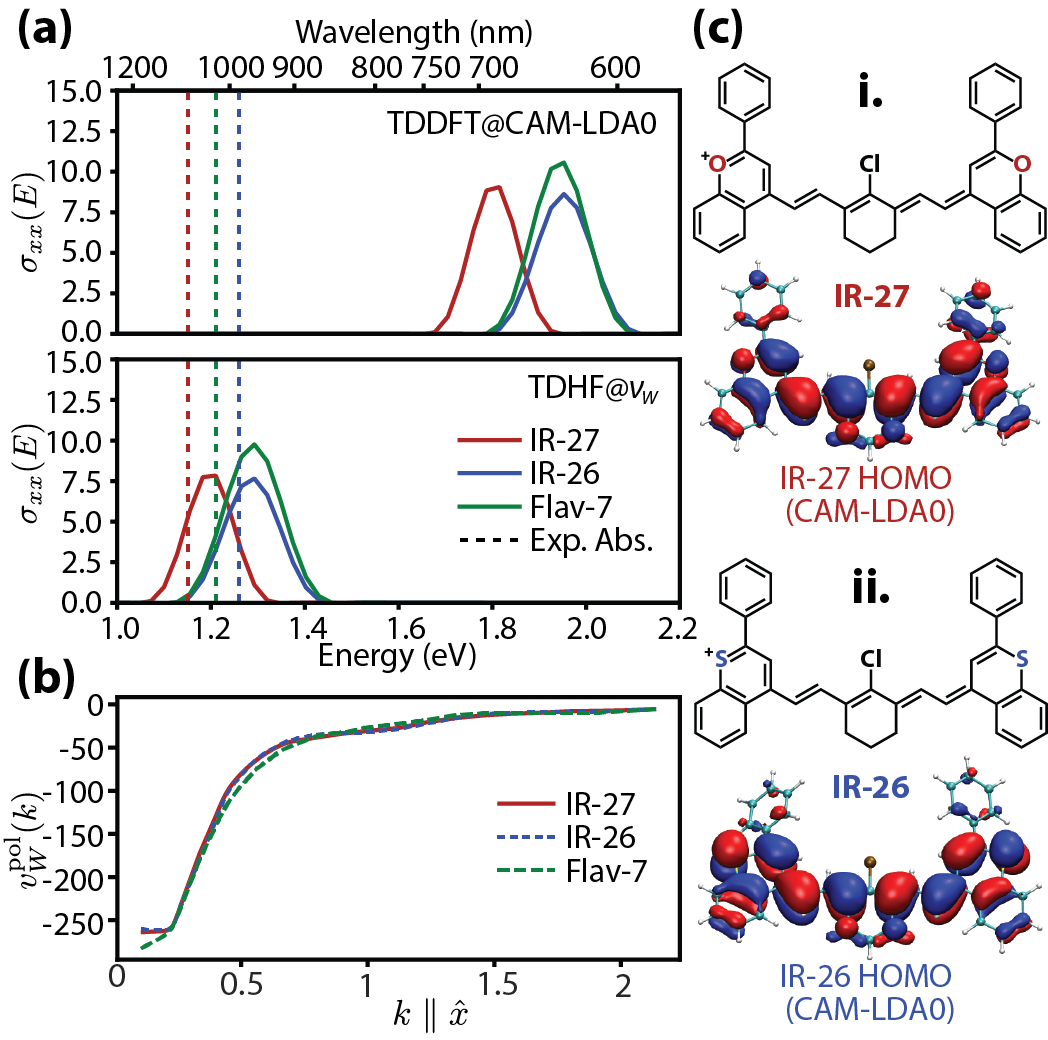}
\caption{\label{fig:ir26}Spectral results (a) and stochastically fitted $v_W$ interaction (b) for the cyanine dyes IR-27 and IR-26 (c).}
\end{figure}

\begin{figure}[b]
\includegraphics{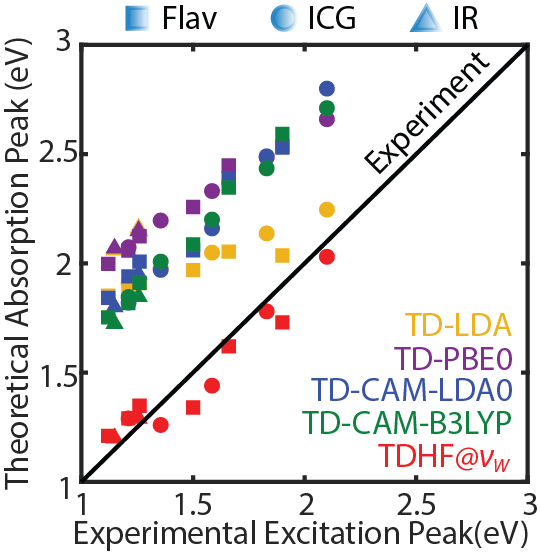}
\caption{\label{fig:graph} Optical gaps from various TDDFT functionals and TDHF@$v_W$ against the water solution experimental absorption peaks.\cite{SeeSI}}
\end{figure}

Figure \ref{fig:wide} presents the principal results of this work for both the Flav (a) and ICG (b) family of dyes. The fitted-interaction TDHF@$v_W$ method significantly enhances the accuracy of predicting the optical gap in polymethine dyes, closely aligning with experimental spectra.\cite{Swamy2024,Langhals2011,Friedman2021,Gamage2024,Heng2022,Cosco2017_ange} For the Flav dyes, the consistent relative error observed between dyes with varying lengths of polymethine bridges in both TDDFT and TDHF@$v_W$ suggests that these errors stem from the ground-state computations and the challenges in optimizing large planar molecules, rather than from inconsistency in the binding energy calculations from  TDHF@$v_W$.

Figure \ref{fig:wide} (b, ii) shows the lowest-energy exciton density of the in ICG-7 extracted directly from the iterative studies,\cite{Neuhauser1990_fdg,bradbury_bethesalpeter_2022,bradbury_optimized_2023} in which the electron is polarized along heterocycle of the dye, giving the expectant strong transition dipole moment of these molecules. Meanwhile, the hole density is concentrated in the $\pi$-bonding orbital of the cyanine backbone and is mostly derived from the HOMO.

Figures \ref{fig:vw} (a) and (b) show the fitted polarization potential, $v_W^{\mathrm{pol}}$, for both dye families, noting their striking similarity. In (c), the ratio between $v_W^{\mathrm{pol}}$ and the bare non-periodic Coulomb interaction $v(k)$ can be thought of as the fraction of exact exchange \textit{removed due to screening} in $k$-space, recalling that the bare Coulomb interaction is $4\pi/k^2$. To facilitate comparison, we also show the amount of exchange removed by screening with the originally optimized parameters of the CAM-LDA0 functional ($\alpha=0.19$, $\beta=0.46$, $\mu=0.33$), and BNL functional with $\gamma=0.15$.\cite{Yanai2004,Tawada2004,Mosquera2016} The fact that the $v_W$ functions for all dyes largely overlap each other shows that the amount of screening (and thereby polarizability) is similar between dyes of varying polymethine bridge length and family (choice of heterocycle). This notably brings hope for a universal screened exchange kernel for use in highly delocalized organic systems. Future studies will determine if such a universal potential exists.

 In (d) and (e), we show the fitted $v_W(k)$ along the cardinal $k$-paths in $k$-space and the corresponding $rv_W(r)$ values in real space, once again showing the fraction of removed exchange due to screening. One observes the nearly isotropic behavior of $v_W$ at high-$k$ (short range) but notable deviations at low-$k$ (long-range). This makes sense as polymethine dye molecules are mostly planar with a carbon bridge extended along one direction ($x$), with a $\pi$ electron density. However, along the $y$ and $z$ directions at long range there is no electron density, and thus reduced screening in $v_W$. We note that in real space it is clear that the CAM-LDA0 and BNL (an optimally tuned long-range exchange functional) are not able to achieve the correct shape nor absolute magnitude of exchange at most length scales, noting especially bad convergence at the short/medium distances.

Figure \ref{fig:ir26} (a) presents results for the Flav-7 dye derivatives, IR-27 and IR-26. We include these dyes as an example where the difference in redshifts is not due to increased binding energy in IR-27, as shown in (b), but is due to a ground-state effect of the LUMO and HOMO energies. This ground-state energy difference is manifested in the difference in HOMO densities near the polymethine bridge due to the electron-withdrawing effect of oxygen (IR-27) versus sulfur (IR-26) (c). Due to how redshifted in energy these dyes are, they serve as another proof that the accuracy of this method does not deteriorate for infrared excitations.  

Figure \ref{fig:graph} compares the optical gaps for all thirteen dyes predicted by various methods with the experimental references.\cite{Swamy2024,Langhals2011,Friedman2021,Gamage2024,Heng2022,Cosco2017_ange} TDHF@$v_W$ achieves a significantly improved mean absolute error (MAE) of 0.09 eV. In contrast, among all TDDFT functionals, TD-CAM-B3LYP produces an MAE of 0.63 eV; TD-CAM-LDA0 yields 0.66 eV;  TD-PBE0 results in 0.79 eV; and TD-LDA shows 0.55 eV. The substantial improvement of TDHF@$v_W$ aligns with the enhancements observed with the full deterministic BSE in model dye systems at a significantly reduced computational cost.\cite{LeGuennic2015} 


Our study has shown the power of a new approach, TDHF@$v_W$, which fits the full many-body screened-exchange interaction based on the BSE within the TDHF framework.\cite{bradbury_bethesalpeter_2022,bradbury_optimized_2023} This method successfully addresses the longstanding challenge of accurately predicting the optical properties of polymethine dye molecules, surpassing the limitations of traditional TDDFT calculations. Analyzing the fitted polarization potential ($v_W^{\mathrm{pol}}$) and the fraction of exact exchange removed due to screening, we identify similarities across different dye families, suggesting the potential for a universal screened exchange kernel applicable to highly delocalized organic systems. Moreover, our findings highlight the consistency of results obtained with TDHF@$v_W$ across various polymethine bridge lengths and heterocycles, offering insights into the fundamental properties of these molecules. Our study also elucidates ground-state effects on the optical properties of dye molecules, as evidenced by differences in redshifts between certain derivatives, which are what most previous TDDFT studied of cyanine dyes detect. This observation reinforces the robustness of TDHF@$v_W$ in capturing complex electronic interactions and ground-state phenomena, particularly for excitations extending into the SWIR region.

Quantitatively, TDHF@$v_W$ exhibits a substantially reduced MAE compared to traditional TDDFT functionals, indicating its superior accuracy in predicting excitation energies. This improvement, achieved at a manageable computational cost, positions TDHF@$v_W$ as a promising tool for researchers in fields such as biological imaging and materials science, where accurate modeling of optical properties is crucial. Additionally, this method realistically scales to bigger sizes, as it is applicable to studying organic chromophore in larger-scale systems, like biological light-harvesting complexes.

\begin{acknowledgments}
    We extend our gratitude to Anthony Spearman and Professor Ellen M. Sletten (UCLA) for their suggestions of polymethine dye molecules. NCB acknowledges the NSF Graduate Research Fellowship Program under grant DGE-2034835. NCB and BYL acknowledge the computing resources provided by ACCESS allocation CHE-230099, CHE-240029, and the Expanse system at the San Diego Supercomputer Center. JRC was supported by NSF grant NSF CHE-2204263. DN was supported by NSF grant CHE-2245253. This work also used computational services associated with the Hoffman2 Shared Cluster provided by UCLA Office of Advanced Research Computing’s Research Technology Group.
\end{acknowledgments}

\bibliography{apssamp}


\end{document}